\documentstyle[12pt]{article}

\def\pasq{q} 
\def\Pasq{Q} 
\def\pas{h}  
\def\Pas{H}  

\def \Plv{Pain\-le\-v\'e}
\def \CRAS{C.~R.~Acad.~Sc.~Paris}
\def \AnnENS{Ann.~\'Ec.~Norm.~}
\def \ie {i.~e.} 
\def \D {\hbox{d}}
\def \Log {\mathop{\rm Log}\nolimits}

\def \pq {\mathop{\rm pq}\nolimits}
\def \ps {\mathop{\rm ps}\nolimits}
\def \qs {\mathop{\rm qs}\nolimits}
\def \rs {\mathop{\rm rs}\nolimits}
\def \discr {\mathop{\rm discr}\nolimits}

\def \mod#1{\vert #1 \vert}

\begin{document}

\pagestyle{plain} 

\begin{center}
 {\bf RULES OF DISCRETIZATION FOR PAINLEV\'E EQUATIONS}
\end{center}

\vskip 0.5 truecm

{\bf Robert Conte\dag} and {\bf Micheline Musette\ddag}

\medskip

\dag
Service de physique de l'\'etat condens\'e,
CEA Saclay,
\hfill \break \indent
F-91191 Gif-sur-Yvette Cedex,
France
\medskip

\ddag
Dienst Theoretische Natuurkunde,
Vrije Universiteit Brussel,
\hfill \break \indent
B-1050 Brussel,
Belgique

\vskip 0.5 truecm

\noindent {\it Short title} :  Discretizing Painlev\'e equations
 
\vglue 1.0truecm
\baselineskip=12truept
 

\vskip 0.8 truecm

\noindent {\it Keywords}
\par discrete Painlev\'e property
\par discrete Painlev\'e test
\par discrete Lax pairs
\par discretization rules
 
\vskip 0.8 truecm

{\it {\bf Abstract} -- 
The discrete Painlev\'e property is precisely defined,
and basic discretization rules to preserve it are stated.
The discrete Painlev\'e test is enriched with a new method which perturbs 
the continuum limit and generates infinitely many no-log conditions.
A general, direct method is provided to search for discrete Lax pairs.
}

\vfill 

\noindent 
{\it Theory of nonlinear special functions : the Painlev\'e transcendents},
eds.~L.~Vinet and P.~Winternitz (Springer, Berlin, 1998).
Montreal, 13--17 May 1996. 
\hfill \break
\noindent

\rightline{\noindent 
           \hskip 1 truemm 17 March 1997
           \hskip 5 truemm Saclay SPEC 96/075 solv-int/9803014}

\eject

\tableofcontents \vfill \eject

\baselineskip=12truept 


\section{Introduction}
\indent

Given some continuous differential equation,
we want to find its discretizations which preserve some global property,
namely the explicit linearizability or more generally
the {\it discrete Painlev\'e property},
a notion recalled section \ref{sectiondPP}
together with its group of invariance.
 From this point of view,
we will never have to address the question of finding a continuum limit,
and consequently we will always write the discrete equations
in a form as close as possible to the canonical form of the continuum limit,
which is well established~\cite{PaiActa,GambierThese}.

Discrete equations are functional equations linking the values taken by some
field variable $u$ at a finite number $N+1$ of points,
either arithmetically consecutive~: $x + k \pas$,
or      geometrically consecutive~: $x \pasq^k,  k-k_0=0,1, \dots, N$,
where $\pas$ or $\pasq$ is the lattice stepsize,
assumed to lie in some neighborhood of, respectively, $0$ or $1$,
and $k_0$ is just some convenient origin.
The integer $N$ is called the {\it order} of the equation,
and we denote, for brevity,
d--(E) and q--(E) these ``difference'' and ``$q-$difference'' equations.
Their study was initiated by Laguerre, 
mainly as three-term ($N=2$) recurrence relations between coefficients of
orthogonal polynomials.
This remained for long a mathematical subject
\cite{Shohat,Freud},
which then extended to topological field theory~\cite{Bessis1979,IZ1980}.
Finally,
the discrete equation
\begin{eqnarray}
& &
E \equiv
-(\overline{u} - 2 u + \underline{u})/ \pas^2
+ 2 (\overline{u} + u + \underline{u}) u + x =0
\label{eqdP1BrezinKazakov}
\end{eqnarray}
already considered by the authors of last five references,
was again encountered by statistical physicists in two-dimensional
quantum gravity~\cite{BrezinKazakov,DouglasShenker,GrossMigdal}
who recognized it as a discrete analogue of the first Painlev\'e equation (P1)
\begin{eqnarray}
& &
E \equiv
- u'' + 6 u^2 + x = 0.
\label{eqP1}
\end{eqnarray}
The same happened simultaneously with a discrete analogue of 
the second Painlev\'e equation (P2)~\cite{PS1990,NP1991}
(in the particular case $\alpha=0$)
\begin{eqnarray}
& &
E \equiv
-(\overline{u} - 2 u + \underline{u}) / \pas^2
+ (\overline{u} + \underline{u}) u^2 + x u + \alpha=0
\label{eqdP2PeriwalShevitz}
\\
& &
E \equiv
- u'' + 2 u^3 + x u + \alpha= 0.
\label{eqP2}
\end{eqnarray}
The above short notation
\begin{eqnarray}
N \hbox{ even}& :  &
                      u  = u(x      ),\
            \overline{u} = u(x +   h),\
           \underline{u} = u(x -   h),\
 \overline{ \overline{u}}= u(x + 2 h),\
\dots
\label{eqNotationEven}
\\
\hskip 7 truemm
N \hbox{ odd}& :  &
            \overline{u} = u(x +   h/2),\
           \underline{u} = u(x -   h/2),\
 \overline{ \overline{u}}= u(x + 3 h/2),\
\dots
\label{eqNotationOdd}
\end{eqnarray}
is adopted throughout the article.

The reason why these two discrete equations,
among many others with the same continuum limit,
deserve the name of discrete Painlev\'e equations, 
in short d--(P1) and d--(P2),
is that they possess the discrete Painlev\'e property.
Indeed, both admit a discrete Lax pair.

Up to now, there exist two methods to find discrete Lax pairs : 
the discrete isomonodromic deformation method
\cite{Birkhoff},
and the discrete analogue~\cite{AL1975}
of the method of Zakharov--Shabat and 
Ablowitz--Kaup--Newell--Segur (AKNS).
But both methods are {\it inverse} methods,
\ie~they generate some discrete equation as a condition betwen the 
coefficients of two given linear operators.
The drawback is that the obtained discrete equation may not be the one
which was looked for.
We propose here a {\it direct} method, based on discretization rules,
to search for the Lax pair of a given discrete equation,
and we obtain several new Lax pairs in this way.

Just like its continuous counterpart,
the discrete Painlev\'e test is the set of all methods one can imagine
to build necessary conditions for a given discrete equation to possess
the discrete Painlev\'e property.
Two such methods are known,
the singularity confinement method~\cite{GRP1991}
and the method of perturbation of the continuum limit~\cite{CM1996}.
The latter is based on the perturbation theorem of Poincar\'e,
which is applicable to differential systems
in an arbitrary number of independent variables,
whether discrete, continuous or even mixed discrete--continuous.
\smallskip

The paper is organized as follows.
The discrete Painlev\'e property (PP) and its group of invariance
are defined in section \ref{sectiondPP},
and a first set of basic rules of discretization is given
in section \ref{sectionBasicRules}.
Section \ref{sectionElliptiqueDiscreteExacte}
recalls as an illustration 
the exact discretizations of the elliptic equations.
In section \ref{sectionLaxDiscrete},
we define discrete Lax pairs from continuous Lax pairs.
Section \ref{sectionLaxDiscretDirect}
states necessary discretization rules for Lax pairs and
details the direct method to obtain the discrete Lax pair of a given equation,
with an application to
the d--(P1) (\ref{eqdP1BrezinKazakov}),
the d--(P2) (\ref{eqdP2PeriwalShevitz})
and a particular d--(P3) of degree one.
In section \ref{sectionMoreLaxPairs},
this method is extended to discrete equations with complementary terms which
do not contribute to the continuum limit,
and several new Lax pairs are obtained.
Section \ref{sectionTestDiscret}
explains the method of perturbation of the continuum limit
for the discrete Painlev\'e test,
and applies it to a qualitative candidate d--(P1).
Finally,
we define in section \ref{sectiondPn}
criteria for a systematic search for the 
fifty discrete Gambier equations.

\section{The discrete Painlev\'e property and its group of invariance}
\label{sectiondPP}
\indent

The (continuous) Painlev\'e property is defined~\cite{PaiLecons}
as the absence of movable critical singularities in the general solution of
a differential equation
\begin{eqnarray}
& &
\forall x\ :\
E(x,u,u',\dots,u^{(N)})=0,
\label{eqContinuous}
\end{eqnarray}
where a singularity is said {\it movable} (as opposed to {\it fixed})
if its location in the complex plane of $x$ depends on the initial conditions,
and {\it critical} if the solution is multivalued around it.
For shortness, following Bureau~\cite{Bureau1939},
we will use the terms ``stability'' for PP, 
``stable'' or ``unstable'' for an equation with or without the PP.

The PP is invariant under the group of birational transformations
\begin{eqnarray}
& &
(u,x) \to (U,X) :\
u=r(x,U,\D U / \D X,\dots,\D^{N-1} U / \D X^{N-1})=0,\
x=\Xi(X),
\\
& &
(U,X) \to (u,x) :\
U=R(X,u,\D u / \D x,\dots,\D^{N-1} u / \D x^{N-1})=0,\
X=\xi(x),\
\label{eqGroupBirationalContinuous}
\end{eqnarray}
($r$ and $R$ rational in $U,u$ and their derivatives,
analytic in $x,X$).
An easier to manage subgroup is made of the homographic transformations
\begin{eqnarray}
& &
(u,x) \to (U,X) :\
u={a U + b \over c U + d},\
X=\xi(x),\
a d - b c \not=0
\label{eqGroupHomographicContinuous}
\end{eqnarray}
where $(a,b,c,d,\xi)$ are arbitrary analytic functions of $x$.
In his classification of second order first degree equations,
Gambier~\cite{GambierThese} has found respectively twenty-four and fifty
 equivalence classes for these two groups 
(with minor later corrections~\cite{BureauMI,CosScou}).

In the discrete case, 
let us consider equations 
\begin{eqnarray}
& &
\forall x\ \forall \pas\ :\
E(x,\pas,\{u(x+k \pas),\ k-k_0=0,\dots,N\})=0
\label{eqDiscretexpas}
\\
& &
\forall x\ \forall \pasq\ :\
E(x,\pasq,\{u(x \pasq^k),\ k-k_0=0,\dots,N\})=0
\label{eqDiscretexpasq}
\end{eqnarray}
algebraic in the values of the field variable,
with coefficients analytic in $x$ and the stepsize $\pas$ or $\pasq$.
It should be noted that $u$ is a function of
{\it two} variables, $x$ and the stepsize.
A natural definition for the discrete Painlev\'e property is the following
\cite{CM1996}.

{\it Definition}.
A discrete equation is said to possess the
{\it discrete Painlev\'e property}
if and only if
there exists a
neighborhood of $\pas=0$ (resp.~$\pasq=1$)
at every point of which
the general solution $x \to u(x,\pas)$ (resp.~$x \to u(x,\pasq)$)
has no movable critical singularities.

{\it Remarks}.
\begin{enumerate}
\item
The definition reduces to that of the continuous PP in the continuum limit.

\item
The singularities in the definition belong to the complex plane of $x$,
not of the stepsize.

\item
This definition immediately extends to equations in an arbitrary number of
independent variables, discrete or continuous,
the extension starting then from the definition of the PP suited to 
partial differential equations (PDEs),
which we do not remind here since this is not our subject.

\end{enumerate}

The discrete PP is invariant under the discrete analogue of
(\ref{eqGroupBirationalContinuous}),
which is the group of nonlocal discrete birational transformations
\begin{eqnarray}
& &
u=r(x,\pas \hbox{ or }\pasq,U,\overline{U},\underline{U},\dots),\
\nonumber
\\
& &
U=R(X,\Pas \hbox{ or }\Pasq,u,\overline{u},\underline{u},\dots),\
X=\xi(x,\pas \hbox{ or }\pasq),\
\Pas =\eta(\pas),\
\Pasq=\kappa(\pasq),\
\label{eqGroupBirationalDiscrete}
\end{eqnarray}
($r$ and $R$ rational in $U,\overline{U},\underline{U},\dots,
u,\overline{u},\underline{u},\dots$,
analytic in $x$ and the stepsize,
$\xi,\eta,\kappa$ analytic).
There exist two discrete analogues of the subgroup
(\ref{eqGroupHomographicContinuous}),
and both may be useful to establish the discrete equivalent of the
classification of Gambier.
The first one is the group of transformations (\ref{eqGroupBirationalDiscrete})
which in the continuum limit reduce to the homographic transformations
(\ref{eqGroupHomographicContinuous}),
where $(a,b,c,d,\xi)$ are arbitrary analytic functions of $x$ and of the
stepsize.
The second one is the group of local homographic transformations
($r$ and $R$ homographic in $U$ and $u$,
independent of 
$\overline{U},\underline{U},\dots,\overline{u},\underline{u},\dots$,
analytic in $x$ and the stepsize,
$\xi,\eta,\kappa$ analytic).

{\it Remarks}.
\begin{enumerate}

\item
The first subgroup seems more useful,
although it does not contain the transformation
$u=\pas^k U,\ k \in {\cal Z}$.

\item
Just like in the continuous case,
the birationality can only be proven by taking the discrete equation into
account.
For instance~\cite{FGR},
given the equation
\begin{eqnarray}
& &
(\overline{u}+u)(u+\underline{u})(4 u+ \pas^4 x - 6) + 8=0,
\end{eqnarray}
and the transformation
\begin{eqnarray}
& &
U(X) =
2/(u(x-\pas/2)+u(x+\pas/2)),
\end{eqnarray}
one first deduces the inverse transformation
\begin{eqnarray}
& &
u(x) =
(6 - 2 U(X - \Pas/2) U(X+ \Pas/2)- \Pas^4 X)/4,\
X=x,\
\Pas=\pas,
\end{eqnarray}
then one plugs it into the direct transformation to get the transformed 
equation
\begin{eqnarray}
& &
(\overline{U}+\underline{U}) U^2 + (\Pas^4 X - 6)U + 4 =0.
\end{eqnarray}
The fields which admit a continuum limit are
$(u-1)/ \pas^2$ and $(1-U)/ \Pas^2$,
this limit being (P1) for both fields.

\end{enumerate}

\section{Basic rules of discretization}
\label{sectionBasicRules}
\indent

Let us now give some basic rules for discretizing a given continuous
equation (\ref{eqContinuous})
into either a difference equation (\ref{eqDiscretexpas})
or a $q$--difference equation (\ref{eqDiscretexpasq}).

The question of discretization is well known in numerical analysis,
where one looks for a {\it scheme} of discretization
which maximizes the order, called {\it scheme order},
of the remainder of the expansion of the left-hand side of
(\ref{eqDiscretexpas})
in a Taylor series of $\pas$ around the center of the $N+1$ points.
A scheme of discretization is said {\it exact} iff it has an infinite order,
like that for the particular (P3) equation 
$\alpha=\beta=\gamma=\delta=0$
\begin{equation}
- u u'' + u'^2 - u u' / x = 0,\
u=c_1 x^{c_2},\
u(x q) u(x /q) - u(x)^2=0.
\end{equation}

Contrary to numerical analysis, only interested in a {\it local} integration,
we require the scheme of discretization to 
preserve the differential order $N$,
an essential element for a {\it global} knowledge of the solution :
every discretization must involve $N+1$ consecutive points.

Like in the continuous case,
there exists another important element concerning discretization rules.

{\it Definition}.
The {\it degree} of a discrete equation is the highest of the two
polynomial degrees of the LHS $E$ of the equation in
$u(x)$ and $u(x + N \pas)$, or $u(x)$ and $u(x {\pasq}^N)$,
where $E$ is assumed polynomial in the $N+1$ variables $u(\dots)$.

The following conjecture has recently been made~\cite{CM1996} :
{\it
``Given an algebraic differential equation with the PP, there exists a 
discretization scheme of order two which conserves the degree.''
}

This conjecture was supported by two examples,
a second order first degree equation with the PP
\begin{eqnarray}
& &
u=(c_1 x + c_2)^2,\
u u'' - (1/2) u'^2 =0,
\end{eqnarray}
and a first order first degree equation without the PP
\begin{equation}
u=(x-c_1)^{-1/2},\
u' + (1/2) u^3 = 0.
\end{equation}
Both admit an exact algebraic discretization,
resulting from the elimination of $c_1$ or $(c_1,c_2)$
between the values of $u$ taken at two or three contiguous points,
and the resulting discrete equation has degree two.
For the first example,
we constructed a first degree discrete equation with the PP,
while we showed the impossibility to do that for the second equation.
Another example is displayed in section \ref{sectiondPn}.

This led us to state the additional rule,
restricted to algebraic differential equations with the PP : 
every second order scheme must also conserve the degree.

\section{The exact discrete elliptic equation
\label{sectionElliptiqueDiscreteExacte}}
\indent

The (continuous) elliptic equation has two usual kinds of normalized forms,
the one of Weierstrass and the twelve ones of Jacobi,
defined by the first and second order equations
($g_2,g_3,a,b$ constants, (p,q,r) arbitrary permutation of (c,d,n))
\begin{eqnarray}
              \wp'^2 &=& 4 \wp^3 - g_2 \wp - g_3
\label{eqWeierstrass1cont}
\\
\wp'' &=& 6 \wp^2 - g_2/2
\label{eqWeierstrass2cont}
\\
            \ps'^2 &=& \qs^2 \rs^2=(\ps^2 + a) (\ps^2 +b)
\label{eqJacobi1cont}
\\
             \ps'' &=& \ps (\qs^2 + \rs^2) = \ps (2 \ps^2 + a + b).
\label{eqJacobi2cont}
\end{eqnarray}
Elliptic functions possess an addition formula,
\ie~an algebraic relation between the values of the function at the
three points $(x_1,x_2,x_1+x_2)$, with $(x_1,x_2)$ arbitrary.
As noticed in the context of discretization
by Baxter~\cite{Baxter1982} and Potts~\cite{Potts1981},
the choice $(x_1,x_2)=(x-\pas /2,\pas)$
{\it ipso facto} defines an exact discretization scheme 
for the first order equations
(\ref{eqWeierstrass1cont}) and (\ref{eqJacobi1cont}).
A scheme for the second order equations
(\ref{eqWeierstrass2cont}) and (\ref{eqJacobi2cont})
then results from the difference of the discrete first order equations
taken for $(x_1,x_2)=(x,\pas)$ and $(x-\pas,\pas)$.
The results are~\cite{Potts1981,Potts1986,Potts1987},
with notation 
(\ref{eqNotationOdd}) for (\ref{eqWeierstrass1disc}) and (\ref{eqJacobi1disc}),
(\ref{eqNotationEven}) for (\ref{eqWeierstrass2disc}) and (\ref{eqJacobi2disc})
\begin{eqnarray}
(\overline{u} - \underline{u})^2 \wp(\pas)
&=& 2 \overline{u} \underline{u}
 (\overline{u} + \underline{u}) - (g_2/2) (\overline{u} + \underline{u}) - g_3
\nonumber
\\
& &
-[(\overline{u} \underline{u} + g_2/4)^2 + g_3 (\overline{u} + \underline{u})]
 \wp^{-1}(\pas)
\label{eqWeierstrass1disc}
\\
(\overline{u} - 2 u + \underline{u}) \wp(\pas)
&=& 2 u (\overline{u} + u + \underline{u}) - g_2/2
\nonumber
\\
& &
- [u^2 (\overline{u} + \underline{u}) + (g_2/2) u + g_3] \wp^{-1}(\pas)
\label{eqWeierstrass2disc}
\\
(\overline{u} - \underline{u})^2 \ps^2 (\pas)
&=&
(\overline{u} \underline{u})^2 
- 2 (\ps' (\pas) + \ps^2 (\pas)) \overline{u} \underline{u}
+ a b
\label{eqJacobi1disc}
\\
(\overline{u} - 2 u + \underline{u}) \ps^2 (\pas)
&=& 
 u^2 (\overline{u} + \underline{u})
- 2 (\ps' (\pas) + \ps^2 (\pas)) u
\label{eqJacobi2disc}
\end{eqnarray}

{\it Remarks}.
\begin{enumerate}
\item
The general solution of (\ref{eqWeierstrass1disc}) and (\ref{eqJacobi1disc})
is by construction
    $\wp(x - x_0,g_2,g_3)$
and $\ps(x - x_0,k)$,
where the step $\pas$ is arbitrary, \ie~{\it not necessarily small},
with $k$ the Jacobi modulus.
These equations therefore possess the discrete PP.
The equations (\ref{eqWeierstrass2disc}) and (\ref{eqJacobi2disc})
also possess the discrete PP since they admit a discrete Lax pair,
see (\ref{eqLaxdwpfullPartiex})--(\ref{eqLaxdwpfullPartiet})
and (\ref{eqLaxdJacPartiex})--(\ref{eqLaxdJacPartiet}).

\item
Order and degree are conserved by the four discretizations.

\item
From these schemes of infinite order,
the Laurent expansion of $\wp(\pas)$ around its double pole $\pas=0$,
or of $\ps(\pas)$ around its simple pole,
defines the second order schemes 
\begin{eqnarray}
(\overline{u} - 2 u + \underline{u}) \pas^{-2}
&=& 2 u (\overline{u} + u + \underline{u}) - g_2/2
\label{eqWeierstrass2discTrunc2}
\\
(\overline{u} - 2 u + \underline{u}) \pas^{-2}
&=& u^2 (\overline{u} + \underline{u}) - (a+b) u,
\label{eqJacobi2discTrunc2}
\end{eqnarray}
which have the discrete PP,
see (\ref{eqLaxdwpshortPartiex})--(\ref{eqLaxdwpshortPartiet})
and (\ref{eqLaxdJacPartiex})--(\ref{eqLaxdJacPartiet}).

\item
Similar equations hold for the nine other Jacobi functions $\pq$,
where the coefficients of the r.h.s.~of (\ref{eqJacobi1disc})
and (\ref{eqJacobi2disc})
are polynomials in $k^2, \pq^2(\pas), \pq'(\pas)$.

\item
Equation (\ref{eqJacobi1disc}) was obtained and integrated in 1973 
by Baxter~\cite{Baxter1982}
in the eight-vertex model,
as a commutation condition of the Yang--Baxter, or star--triangle relations.
These Yang--Baxter relations~\cite{JimboYangBaxter},
which are second order discrete tensorial equations,
play in the discrete domain a role as central as the one played by
the Yang--Mills equations in the continuous domain.

\end{enumerate}

\section{Discrete Lax pairs}
\label{sectionLaxDiscrete}
\indent

In the two relations defining a Lax pair of a (continuous) ODE
$E(x,u)=0$,
e.g.~in matrix form with $t$ the spectral parameter
\begin{eqnarray}
& &
\partial_x \psi = L \psi,\
\partial_t \psi = M \psi,\
(C \equiv \partial_t L - \partial_x M + L M - M L =0) \Leftrightarrow (E=0)
\end{eqnarray}
one can discretize either $x$ alone or $x$ and $t$.
Let us restrict here to the first case and to the difference type equations.

The rule of conservation of the differential order
requires the column vector $\psi$ to be discretized with two points,
which we denote
$\overline{\psi}=\psi(x + \pas /2)$ and $\underline{\psi}=\psi(x - \pas /2)$,
and $L$ to be discretized with as many points $u$ as required by the
differential order of the equation under consideration,
points which we denote
$\overline{u}=u(x + \pas), u=u(x), \underline{u}=u(x- \pas)$
for a second order equation.
In order to keep a linear correspondence between the continuous operators
$(L,M)$ and their discrete counterparts,
it is then convenient to discretize $\partial_x \psi = L \psi$
in a dissymmetric-looking way and to introduce~\cite{Birkhoff} 
the linear operator $A$ linking $\underline{\psi}$ to $\overline{\psi}$,
thus defining the discrete Lax pair
$(A,B,z,\psi,\pas)$ as~\cite{IKF1990}
\begin{eqnarray}
& &
            \overline{\psi} = A \underline{\psi},\
\partial_z \underline{\psi} = B \underline{\psi},\
(K \equiv \partial_z A + A B - \overline{B} A =0) \Leftrightarrow (E=0).
\label{eqLaxDiscretCommutateur}
\end{eqnarray}
The continuum limit is then
\begin{eqnarray}
& &
{A - 1 \over \pas} \to L,\
(\D z / \D t) B \to M,\
\nonumber
\\
& &
(\D z / \D t) (\partial_z A + A B - \overline{B} A) / \pas
\to 
\partial_t L - \partial_x M + L M - M L,
\label{eqLaxLimiteContinue}
\end{eqnarray}
with some link $F(t,z,\pas)=0$ between the spectral parameters $t$ and $z$.

For a second order equation $E(\overline{u},u,\underline{u},x,\pas)=0$,
the operators $A$ and $B$ must have the $u-$dependences 
$A(\overline{u},u,\underline{u}),B(u,\underline{u})$.

{\it Remark}.
The definition is invariant under the involution
\begin{eqnarray}
& &
(E,A     ,B,x,  \pas,\overline{u},u,\underline{u}) \to
(E,A^{-1},B,x,- \pas,\underline{u},u,\overline{u}),
\label{eqLaxSymetrie}
\end{eqnarray}
which sometimes allows to suppress an undesired denominator in matrix $A$,
like in the matricial Lax pair of d--(P1) given in Ref.~\cite{JBH1992}.

The interest of a discrete Lax pair is to provide a constructive proof of the
PP,
just like in the continuous case.

\section{A direct method towards matricial Lax pairs}
\label{sectionLaxDiscretDirect}
\indent

The two methods recalled in the introduction to find discrete Lax pairs
are inverse methods.
 From the point of view of discretization,
a new, direct method emerges, which is as follows.

Let be given 
a continuous equation, its Lax pair,
and some discretization of the continuous equation.
We first state general rules for discretizing the Lax pair,
involving some free functions;
we then enforce the cross-derivative condition
(\ref{eqLaxDiscretCommutateur})
to remove all the freedom on these functions,
except the link between $t$ and $z$ which cannot be removed;
finally,
we choose the link between $t$ and $z$,
so as to preserve the rational dependence of the matricial Lax pair on the
spectral parameter
and to ensure the existence of the continuum limit of the discrete Lax pair.

In this section,
we handle 
the d--(P1) (\ref{eqdP1BrezinKazakov})
and
the d--(P2) (\ref{eqdP2PeriwalShevitz}),
whose matricial Lax pairs with a continuum limit have been found
respectively by Refs.~\cite{JBH1992,GNPRS1994}, using an inverse method,
and by Ref.~\cite{CM1996}, using the present method.
We also handle the d--(P3) of Ref.~\cite{GNPRS1994} when its degree two
reduces to one, i.e.~for $(\alpha,\beta,\gamma,\delta)=(0,0,0,0)$
\begin{eqnarray}
& &
E \equiv
-x u (\overline{u} - 2 u + \underline{u})/ \pas^2
+x (\overline{u} - u) (u - \underline{u})/ \pas^2
-u (\overline{u} - \underline{u})/ (2 \pas) =0,
\label{eqdP3}
\end{eqnarray}
with continuum limit either of the two canonical forms of the third Painlev\'e
equation, (P3) or (P3') (Ref.~\cite{PaiCRAS1906} p.~1115)
\begin{eqnarray}
({\rm P3'})\ 
E
&\equiv&
-u''+{u'^2 \over u} - {u' \over x} + {\alpha u^2 + \gamma u^3 \over 4 x^2}
 + {\beta  \over 4 x}
 + {\delta \over 4 u}=0,\ 
\label{eqP3prime}
\\
({\rm P3})\ 
E
&\equiv&
-u''+{u'^2 \over u} - {u' \over x} + {\alpha u^2 + \beta \over x}
 + \gamma u^3 + {\delta \over u}=0.
\label{eqP3}
\end{eqnarray}

One must first make a choice between three kinds of Lax pairs for the (Pn) 
equations :
the second order scalar ``Lax'' pairs of Garnier~\cite{GarnierThese},
the second order matricial ones of Jimbo and Miwa~\cite{JimboMiwaII},
and the ones of Flaschka and Newell~\cite{FlaschkaNewell}
arising from the reduction of a PDE.
In this paper,
we restrict to the third type in matricial form when the matrix order is two.
Using Pauli matrices
\begin{eqnarray}
& &
\sigma_1=\pmatrix{0 & 1 \cr 1 & 0 \cr },\ 
\sigma_2=\pmatrix{0 &-i \cr i & 0 \cr },\ 
\sigma_3=\pmatrix{1 & 0 \cr 0 &-1 \cr },\ 
\sigma_j \sigma_k= \delta_{jk} + i \varepsilon_{jkl} \sigma_l,\
\label{eqPauli}
\\
& &
\sigma^{+}=\pmatrix{0 & 1 \cr 0 & 0 \cr },\ 
\sigma^{-}=\pmatrix{0 & 0 \cr 1 & 0 \cr },\ 
\nonumber
\end{eqnarray}
the continuous pairs are as follows. 

For (P1) \cite{JimboMiwaII} :
\begin{eqnarray}
& &
L=2 (u - t) \sigma^{+} + \sigma^{-},\ 
\label{eqLaxP1partiex}
\\
& &
M=2 u' \sigma_3 +(-4 u^2-2 x +8 t u -16 t^2)\sigma^{+}+(4 u + 8 t)\sigma^{-},\
C=2 \sigma_3 E,
\label{eqLaxP1partiet}
\end{eqnarray}

For (P2) \cite{FlaschkaNewell} :
\begin{eqnarray}
& &
L= - t \sigma_3 + u \sigma_1,\
\label{eqLaxP2partiex}
\\
& &
M=(4 t^2-2 u^2 - x)\sigma_3+2 u' i \sigma_2 -(4 t u + \alpha t^{-1})\sigma_1,\
C= 2 i \sigma_2 E,
\label{eqLaxP2partiet}
\end{eqnarray}

For the (P3) (\ref{eqP3}) :
\begin{eqnarray}
L &= &
(1/2)(u'/u + c u + d/u) \sigma_3 + t \sigma_1,\
c^2= \gamma, d^2=- \delta,
\label{eqLaxP3partiex}
\\
M &= &
\Big[
(x/(2t))(u'/u + c u + d/u) \sigma_3 + x \sigma_1
+t^{-2} c d x \sigma_1/2
-t^{-3} (\alpha c - \beta d) \sigma_3
\nonumber
\\
& &
-t^{-2} (\alpha u + \gamma x u^2 + c (x    u'  +   u)) \sigma^{+}/2
\nonumber
\\
& &
-t^{-2} (\beta/ u + \delta x/u^2 + d (x (1/u)' + 1/u)) \sigma^{-}/2
\Big]  {t^2 \over t^2 + c d}
\label{eqLaxP3partiet}
\\
C &= &
((x/(2 t u)) (\sigma_3 -(c u /t) \sigma^{+} +(d/(t u)) \sigma^{-}) 
{t^2 \over t^2 + c d} E.
\end{eqnarray}
This Lax pair for (P3) is the extension to arbitrary 
$\alpha,\beta,\gamma,\delta$
of the pair given by Milne~\cite{MilneThesis} for $\gamma \delta=0$.

The rules of discretization are the following.

\begin{enumerate}
\item
conserve the matricial order.
This is indeed the differential order of the scalar Lax pair,
which must be conserved;

\item
replace the continuous spectral parameter $t$ by an unspecified
function $T(z,\pas)$;

\item
discretize the operator $L$ centered at the three points $x-\pas,x,x+\pas$.
If $L$ is traceless,
so is its discretization;

\item
discretize the operator $M$ centered at the two points $x-\pas,x$.
If $M$ is traceless, so is its discretization;

\item
for each matrix element,
enforce conservation of order and degree;

\item
replace each monomial $(\D u / \D x)^k$ by its discretization obeying
the general rules,
multiplied by the $k$-th power of an unspecified function $g(z,\pas)$.
This function $g$,
whose continuum limit must be $1$ for any $z$,
represents the ratio of the stepsize $\pas$ to the differential element $\D x$;
\item
take $B$ as the product of the discretized $M$ by an unspecified function
$J(z,\pas)$
(like Jacobian)
representing a discretization of the derivative $\D T / \D z$;

\item
take $(A-1)/ \pas$ equal to the sum of the discretized operator $L$
and a diagonal matrix of unspecified functions of $(z,\pas)$ only,
diag$(g_1,g_2)$;
these functions,
whose continuum limit must be zero,
account for the dissymmetry of the formula defining $A$.

\end{enumerate}

In the second order, first degree case of Painlev\'e equations (Pn),
examples of discretizations obeying the above rules are
\begin{eqnarray}
u^2 \hbox{ in } L &\to&
\nu_1 u (\overline{u} + \underline{u})/2
+\nu_2 u^2+\nu_3 \overline{u}\underline{u},\
\nu_1+\nu_2+\nu_3=1,
\nonumber
\\
u^2 \hbox{ in } M &\to&
u \underline{u}
\nonumber
\\
x \hbox{ in } M &\to&
x - \pas/2.
\end{eqnarray}

Finally,
after the cross-derivative condition has been enforced,
one must perform the continuum limit on those of the functions 
$T,g,J,g_1,g_2$ which remain unspecified.
One such free function is $T$,
because the choice of $z$ is arbitrary.
One chooses $T(z,\pas)$ as a rational function of $z$
(to conserve the rational dependence of the Lax pair on its spectral
parameter)
such that the inverse function $z$ of $(t,\pas)$
admits for every $t$ a finite nonzero limit when $\pas \to 0$.

The discretized matricial Lax pair of 
the d--(P1) (\ref{eqdP1BrezinKazakov}) is, with $\lambda_1 + \lambda_2 =1$
\begin{eqnarray}
A&=& 1 + \pas
 \pmatrix{g_1 & 2\lambda_1 u + \lambda_2 (\overline{u} + \underline{u})
 - 2 T  \cr 1 & g_2 \cr},\
\nonumber
\\
B/J&=& \pmatrix{
  2 g (u - \underline{u}) / \pas &
 -4 u \underline{u} -2(x - \pas /2) + 4 T (u + \underline{u}) - 16 T^2 \cr
  2 (u + \underline{u}) + 8 T &
 -2 g (u - \underline{u}) / \pas \cr},
\label{eqLaxdP1Assump}
\end{eqnarray}
and that of
the d--(P2) (\ref{eqdP2PeriwalShevitz}) is
\begin{eqnarray}
A&=&1 + \pas (-T \sigma_3 + 
(\lambda_1 u + \lambda_2 (\overline{u} + \underline{u})/2) \sigma_1)
+ \pas \pmatrix {g_1 & 0 \cr 0 & g_2 \cr}),\
\nonumber
\\
B/J&=&
(-2 u \underline{u} + 4 T^2 - (x - \pas /2)) \sigma_3
+ 2 g ((u-\underline{u})/ \pas) (i \sigma_2)
\nonumber
\\
& &
- (2 T (u+\underline{u}) + \alpha/T) \sigma_1
\end{eqnarray}
depending on the yet unspecified functions $T,g,J,g_1,g_2$ of $(z,\pas)$.
The cross-derivative condition is enforced by
eliminating
for instance $\overline{u}$ 
(or here the variable $x$, which appears always at the first power)
between the condition $K=0$ 
(\ref{eqLaxDiscretCommutateur})
and the discrete equation,
and identifying to zero the resulting polynomial in the three variables 
$(x,u,\underline{u})$
(or here the three variables $(\overline{u},u,\underline{u})$).
This results in the relations
\begin{eqnarray}
\hbox{d--(P1)} & &
\lambda_1=1,\
g_1=g_2=(g-1)/ \pas,\
1 - g^2 - 2 \pas^2 T^2=0,\
J=-g' / \pas^2,\
\\
\hbox{d--(P2)} & &
\lambda_1=1,\
g_1=g_2=(g-1)/ \pas,\
1 - g^2 + \pas^2 T^2=0,\
J=g' / (\pas^2 T),\
\end{eqnarray}
and one of the five functions remains free, as expected.
A convenient choice of $T$ is
\begin{eqnarray}
\hbox{d--(P1)} & &
T=z - \pas^2 z^2/2,\
z=t + (t^2/2) \pas^2 + O(\pas^4),
\\
\hbox{d--(P2)} & &
T={z - z^{-1} \over 2 \pas},\
z=1 + \pas t + O(\pas^2).
\label{eqdP2convenient}
\end{eqnarray}

Finally, the matricial Lax pair of
the d--(P1) (\ref{eqdP1BrezinKazakov})
is
\begin{eqnarray}
A&=&\pmatrix{1-\pas^2 z & \pas (2 u - 2 t) \cr \pas & 1-\pas^2 z \cr},\
t=z - \pas^2 z^2/2,\
z=t + (t^2/2) \pas^2 + O(\pas^4)
\nonumber
\\
B&=& \pmatrix{
  2 (1 - \pas^2 z) (u - \underline{u}) / \pas &
-4 u \underline{u} -2(x - \pas /2) + 4 t (u + \underline{u}) - 16 t^2 \cr
2 (u + \underline{u}) + 8 t &
 -2 (1 - \pas^2 z) (u - \underline{u}) / \pas \cr},
\label{eqLaxdP1BrezinKazakov}
\\
\pas^{-1} K &=& 2 \sigma_3 E,
\nonumber
\end{eqnarray}
a result obtained~\cite{CM1996} with this method.
This is the rewriting of eqns.~(1.12)--(1.13) of Ref.~\cite{FGR}
which clearly shows the continuum limit,
under the change of basis
\begin{eqnarray}
P&=&
\pmatrix{f & 0 \cr 0 & \underline{f} \cr}
\pmatrix{1 & \zeta \cr 0 & 1 \cr},\
\left(\overline{P}\right)^{-1} A P =
\pas \pmatrix{2 \zeta/ \sqrt{\overline{\omega}} & 
              - \sqrt{\omega / \overline{\omega}}\cr 1 & 0 \cr},\
\\
& &
\zeta= (1 - \pas^2 z) / \pas,\
\omega= (1 - 2 \pas^2 u) / \pas^2,\
\underline{f} / f = \sqrt{1 - 2 \pas^2 u}.
\nonumber
\end{eqnarray}

As to the matricial Lax pair of
the d--(P2) (\ref{eqdP2PeriwalShevitz})
{
\begin{eqnarray}
A&=& \pmatrix {1/z & \pas u \cr \pas u & z \cr},\
t={z - z^{-1} \over 2 \pas},\
z=1 + \pas t + O(\pas^2)
\nonumber
\\
\pas z B&=&
(-2 u \underline{u} + 4 t^2 - (x - \pas /2)) \sigma_3
+ (z+1/z) ((u-\underline{u})/ \pas) (i \sigma_2)
\nonumber
\\
& &
- (2 t (u+\underline{u}) + \alpha/t) \sigma_1
\label{eqLaxdP2PeriwalShevitz}
\\
z K &=& 2 i \sigma_2 E.
\nonumber
\end{eqnarray}
}
it was already known~\cite{JBH1992,GNPRS1994}. 

The case of d--(P3) is slightly more difficult due to denominators
$u$ and $x$ in the expression of $u''$,
so the discretization of $u'/u$ in $L$ may be anything in between
$\discr(u')/ \discr(u)$ and $\discr(x u u') / \discr(x u^2)$,
where $\discr()$ symbolizes a discretization obeying the rules,
and the same also applies to the term $x u'/u$ in $M$.
This difficulty is overcome by noticing that the only term of the commutator 
$C$ to generate the equation is $M_x$,
due to the first integral $x u'/u$ of this very particular (P3).
This first integral has the discrete analogue
$f(x)=(2/\pas) x (u(x+\pas/2) - u(x-\pas/2)) / (u(x+\pas/2) + u(x-\pas/2))$,
so the discretization of $x u' / u$ in $M$ can only be $g(z) f(x-\pas/2)$
according to our rules.
We therefore start from the assumption
\begin{eqnarray}
A&=&\pmatrix {G_1 & 0 \cr 0 & G_2 \cr}
 + \pas (g F(x) \sigma_3/2 + T \sigma_1),\
\nonumber
\\
B/J&=&
g (f(x-\pas/2) + f_0) \sigma_3/(2 T) + (x-\pas/2) \sigma_1,\
\end{eqnarray}
with six unknown functions of $z$ ($T,g,J,G_1,G_2,f_0$)
and one unknown function of $x$ ($F$)
representing the discretization of $u'/u$.
The condition that the discrete equation be a factor of the commutator $K$
(\ref{eqLaxDiscretCommutateur})
yields
\begin{eqnarray}
\hbox{d--(P3)} & &
G_1=G_2,\
g'=0,\
f_0'=0,\
J=G_1' / (\pas^2 T),\
(G_1^2  - \pas^2 T^2)'=0,\
\nonumber
\\
\phantom{\hbox{d--(P3)}} & &
F(x)=(f(x+\pas/2)+f(x-\pas/2)+ 2 K_0)/(2 x),\
K_0 \hbox{  constant},
\end{eqnarray}
resulting in the same convenient choice of $T(z)$ as in d--(P2),
eq.~(\ref{eqdP2convenient}).

The resulting Lax pair for this particular d--(P3) is finally
\begin{eqnarray}
f(x)&=&
x {u(x+\pas/2)-u(x-\pas/2) \over u(x+\pas/2)+u(x-\pas/2)} {2\over \pas},\
F(x)={f(x+\pas/2)+f(x-\pas/2)+ 2 K_0 \over 2 x},\
\nonumber
\\
A&=&(z+z^{-1})/2 + \pas (F(x) \sigma_3/2 + ((z-z^{-1})/(2 \pas)) \sigma_1),\
\nonumber
\\
\pas z B&=&
{\pas \over z-z^{-1}} f(x-\pas/2) \sigma_3 + (x-\pas/2) \sigma_1,\
\nonumber
\\
z K &=& {h \over z - z^{-1}} ((z+z^{-1}) \sigma_3/2 + \pas f(x) \sigma_1) E.
\label{eqLaxdP3Begin}
\end{eqnarray}

{\it Remarks}.
\begin{enumerate}

\item
The scalar Lax pair satisfied by the second 
component $\psi_2$ of $\psi$ in
(\ref{eqLaxdP1BrezinKazakov})
\begin{eqnarray}
& &
(\overline{\psi_2} -2 \psi_2 + \underline{\psi_2}) / \pas^2 
- 2 \underline{u} \underline{\psi_2} + 2 z \psi_2 =0,\
\nonumber
\\
& &
\psi_{2,z}
 + 2 [4 z (\psi_2 - \overline{\psi_2} + 2 u \psi_2 - u \overline{\psi_2}
 - \underline{u} \overline{\psi_2}] / \pas
\nonumber
\\
& &
\phantom{\psi_{2,z}}
+ 4 [z^2 (\overline{\psi_2} - 3 \psi_2) - z u \psi_2] \pas
+ 4 z^3 \psi_2 \pas^3
=0
\label{eqLaxdP1Scalar}
\end{eqnarray}
is of course the one obtained by Joshi {\it et al.}~\cite{JBH1992};
as to the scalar Lax pair found in quantum gravity
\cite{IZ1980,IKF1990},
which contains algebraic coefficients,
it is the transformed of this one by $\psi_3= G \psi_2$,
with $G$ satisfying the discrete equation
$G / \underline{G} = \sqrt{1 - 2 \pas^2 u}$.

\item
The term $u'/u$ of operator $L$ for (P3) is discretized as the quotient 
of $\discr(x u u')$ by $\discr(x u^2)$,
a result not easy to guess in advance.

\item
The number of singular points in the complex plane of the spectral parameter
is not necessarily the same for the matrix $M(t)$ and for its discretized
$B(z)$.
For (P2), $M$ has two singular points $t=0,\infty$
while $B$ has four such points $z=0,\infty,1,-1$.

\end{enumerate}

One similarly obtains a Lax pair 
for the discrete Weierstrass equation
without complementary terms
(\ref{eqWeierstrass2discTrunc2})
\def\Z{\lambda}
\begin{eqnarray}
t&=& (1-\Z^2)/ (2 \pas^2),\
\nonumber
\\
A&=&
\pmatrix{\Z & 2 \pas(u - t) \cr \pas & \Z \cr},\
\label{eqLaxdwpshortPartiex}
\\
B&=& 2 \Z ((u - \underline{u})/ \pas)  \sigma_3
+ (-16 t^2 + 4 t (u + \underline{u}) + g_2 - 4 u \underline{u}) \sigma^{+}
\nonumber
\\
& &
+             (2 (u + \underline{u}) + 8 t) \sigma^{-}
\label{eqLaxdwpshortPartiet}
\\
\pas^{-1} K &=& 2 \sigma_3 E
\nonumber
\end{eqnarray}
and one for the discrete Jacobi equation (\ref{eqJacobi2discTrunc2})
\begin{eqnarray}
t&=& (\Z - 1/ \Z) / (2 \pas),
\nonumber
\\
A&=& \pmatrix {1/\Z & \pas u \cr \pas u & \Z \cr},\
\label{eqLaxdJacPartiex}
\\
B&=& 
(-2 u \underline{u} + 4 t^2 - a-b)       \sigma_3
+ (\Z+1/\Z) ((u-\underline{u})/ \pas) (i \sigma_2)
\nonumber
\\
& &
- 2 t (u+\underline{u})                  \sigma_1
\label{eqLaxdJacPartiet}
\\
\pas^{-1} K &=& 2 i \sigma_2 E.
\nonumber
\end{eqnarray}

These last two Lax pairs do not depend on the spectral parameter $z$,
and $\Z$ is an arbitrary constant.

\section{More on discrete Lax pairs}
\label{sectionMoreLaxPairs}
\indent

Some discrete equations have complementary terms which do not contribute to
the continuum limit,
such as 
the discrete Weierstrass equation
with its terms $\wp^{-2}(h)$,
or 
the following d--(P1) (Ref.~\cite{FGR} eq.~(2.8)),
which only differs from the d--(P1) (\ref{eqdP1BrezinKazakov}) by
terms homogeneous to $\pas^2 x u$ and $\pas^2 u^3$
\begin{eqnarray}
E & \equiv &
-(\overline{u} - 2 u + \underline{u})/ \pas^2
+ 2 (\overline{u} + u + \underline{u}) u + x
\nonumber
\\
& &
- \pas^2 x u - \pas^2 u^2 (\overline{u} + u + \underline{u}) =0.
\label{dP1ordre4}
\end{eqnarray}

A Lax pair has already been obtained~\cite{FGR},
not for this d--(P1) exactly,
but only for the discrete derivative of a birational transform of it,
which makes it not simple at all.
Let us obtain the natural Lax pair.

One assumes the same qualitative form (\ref{eqLaxdP1Assump})
than for the d--(P1) without complementary terms,
and one adds to $A$ and $B$ as many matrices of order at least $\pas$
as there exist divisors of these complementary terms,
each matrix being the product of such a divisor
by a matrix of free functions of $z$.
The result for (\ref{dP1ordre4}) is
\begin{eqnarray}
t&=&(1-z)/ \pas^2,\
z=1 - \pas^2 t,\
\nonumber
\\
A&=&
\pmatrix{z & (2/ \pas)(z-1+ \pas^2 u) \cr (\pas/2)(z+1- \pas^2 u) & z \cr},\
\label{eqLaxdP1fullPartiex}
\\
- \pas^2 z B&=&
 2 z ((u - \underline{u})/ \pas)      \sigma_3
- \pas (2 u \underline{u} +x - \pas/2)
                                    \pmatrix{0 & 2 / \pas \cr \pas /2 & 0 \cr}
\nonumber
\\
& &
+(u + \underline{u})                \pmatrix{0 & 4 t \cr  1+z & 0 \cr}
-2 t (1+z)/z                        \pmatrix{0 & 4 t \cr -1-z & 0 \cr}
\label{eqLaxdP1fullPartiet}
\\
\pas^{-1} K &=& 2 \sigma_3 E.
\nonumber
\end{eqnarray}
This Lax pair admits by construction 
(\ref{eqLaxP1partiex})--(\ref{eqLaxP1partiet})
as its continuum limit.

One similarly obtains a Lax pair for the discrete Weierstrass equation 
with complementary terms
(\ref{eqWeierstrass2disc})
\begin{eqnarray}
t&=& (1-\Z)/ \Pas^2,\
\Pas^{-2} = \wp(\pas),\
\nonumber
\\
A&=&
\pmatrix{\Z & (2/ \Pas)(\Z-1+\Pas^2 u) \cr (\Pas/2)(\Z+1- \Pas^2 u) & \Z \cr},
\label{eqLaxdwpfullPartiex}
\\
B&=& 2 \Z ((u - \underline{u})/ \Pas)  \sigma_3
- \Pas (2 u \underline{u} - g_2/2 - \Pas^2 g_3)
                                    \pmatrix{0 & 2 / \Pas \cr \Pas /2 & 0 \cr}
\nonumber
\\
& &
+(u + \underline{u})                \pmatrix{0 & 4 t \cr  1+\Z & 0 \cr}
\label{eqLaxdwpfullPartiet}
\\
& &
-2 t ((1+\Z)/\Z -(\Pas^4 g_2 + \Pas^6 g_3)/4)
                                    \pmatrix{0 & 4 t \cr -1-\Z & 0 \cr}
\nonumber
\\
\Pas^{-1} K &=& 2 \sigma_3 E.
\nonumber
\end{eqnarray}

\section{The discrete Painlev\'e test
\label{sectionTestDiscret}}
\indent

In the continuous case,
{\it all} the methods of the Painlev\'e test,
without exception,
are based on two theorems and only two,
namely
the existence theorem of Cauchy
and the theorem of perturbations of Poincar\'e~\cite{Poincare}.
This is explained in detail in
the lecture notes of a Chamonix school~\cite{Chamonix1993},
an updated version of which is in preparation~\cite{Cargese1996}.

For discrete equations,
this is also the case for all methods but one,
the singularity confinement method,
which really seems outside the scope of the theorem of Poincar\'e.

Consider an arbitrary discrete equation (\ref{eqDiscretexpas}),
also depending on some parameters $a$,
and 
let $(x, \pas, u, a) \to (X, \Pas, U, A, \varepsilon)$
be an arbitrary perturbation admissible by the theorem of Poincar\'e
(which excludes any nonanalyticity,
like $\varepsilon^{1/5}$, for the perturbed variables).
A necessary condition is that the limit $\varepsilon=0$ possesses the PP
(discrete or continuous, this does not matter).

To illustrate the different methods, 
let us discretize the equation (P1) by a second order scheme,
using the rules previously stated
\begin{eqnarray}
E & \equiv & 
- (\overline{u} - 2 u + \underline{u}) \pas^{-2} 
+ 3 \lambda_1 (\overline{u} + \underline{u}) u
+ 6 \lambda_2 u^2
+ 6 \lambda_3 \overline{u} \underline{u}
+ g=0.
\label{eqP1CompleteDiscrete}
\end{eqnarray}
with $\sum \lambda_k=1$,
and $g$ an unspecified function of $x$.

The test will generate necessary conditions on
$(\lambda_k, g)$.
One must find at least the following solution,
where the equation has a Lax pair :
$g=x$ and 
$\overrightarrow \lambda=(2/3,1/3,0)$;
one may also find the following solutions,
isolated by the singularity confinement method :
$g=x$ and 
$\overrightarrow \lambda=(1,0,0)$       (Ref.~\cite{GRP1991}),
$\overrightarrow \lambda=(1/2,1/4,1/4)$ (Ref.~\cite{RG96} eq.~(5.5)).

\subsection{Singularity confinement method}
\label{sectionConfinement}
\indent

If the field $u$ admits a pole at some point $x_0$ in the complex plane $x$
\begin{eqnarray}
& &
u(x) \sim u_0 \chi^p,\ \chi=x-x_0 \to 0,\ u_0 \not=0,\ -p \in {\cal N},
\end{eqnarray}
it is generically regular at any point $x_0+x_1$
where $x_1$ is not infinitesimal
\begin{eqnarray}
& &
\forall x_1,\ \mod{x_1} >> 0 \ : \
u(x_0+x_1) \not= \infty.
\label{eqConfinementCondition}
\end{eqnarray}
When $u$ satisfies a discrete equation of order $N$, the implementation of
this ``confinement condition''
consists in requiring the property (\ref{eqConfinementCondition})
for $N+1$ consecutive iterates,
which generically ensures the property for the next iterates.
The polar behaviour is then only sensitive during a finite number of
iterations.

\subsection{Method of perturbation of the continuum limit
\label{sectionMethodePerturbativeLimiteContinue}}
\indent

Defined by an expansion of $u$ as a Taylor series in the lattice stepsize
\cite{CM1996}
\begin{eqnarray}
& &
x      \hbox{ unchanged},\
\pas = \varepsilon,\
\pasq = e^{\varepsilon},\
u    = \sum_{n=0}^{+ \infty} \varepsilon^n u^{(n)},\
a    = \hbox{ analytic }(A, \varepsilon),
\label{eqPerturbationLimiteContinue}
\end{eqnarray}
this perturbation generates an infinite sequence of differential equations 
$E^{(n)}=0$
\begin{eqnarray}
& &
E = \sum_{n=0}^{+ \infty} \varepsilon^n E^{(n)}
\\
& &
E^{(n)} (x, u^{(0)}, \dots, u^{(n)})
\equiv
 {E^{(0)}(x, u^{(0)})}' u^{(n)} + R^{(n)}(x, u^{(0)}, \dots, u^{(n-1)}) = 0,\ 
n\ge 1,
\end{eqnarray}
whose first one $n=0$ is the ``continuum limit''.
The next ones $n\ge 1$, which are linear inhomogeneous,
have the same homogeneous part $E^{(0)'} u^{(n)}=0$ independent of $n$,
defined by the derivative of the equation of the continuum limit,
while their inhomogeneous part $R^{(n)}$ (``right-hand side'')
comes at the same time from the nonlinearities and the discretization.

This perturbation of the continuum limit is entirely analogous to the
perturbative method of the continuous case,
either in its Fuchsian version~\cite{CFP1993} or 
in its nonFuchsian one~\cite{MC1995},
depending on the nature, Fuchsian or nonFuchsian,
of the linearized equation $E^{(1)}=0$ at a singulier point of $u^{(0)}$.

The simplicity of the method is best seen on the Euler scheme for the 
Bernoulli equation~\cite{CM1996} 
\begin{equation}
E \equiv (u(x+h) - u(x)) / \pas + u(x)^2 = 0,\
\label{eqdRiccatiEuler}
\end{equation}
i.e.~the logistic map of Verhulst,
a paradigm of chaotic behaviour which should therefore fail the test.
Let us expand the terms of (\ref{eqdRiccatiEuler})
according to the perturbation (\ref{eqPerturbationLimiteContinue})
up to an order in $\varepsilon$ sufficient to build the first 
equation $E^{(1)}=0$ beyond the continuum limit $E^{(0)}=0$
\begin{eqnarray}
u & = &
u^{(0)} + u^{(1)} \varepsilon + O(\varepsilon^2)
\\
u^2 & = &
u^{(0)^2}
 + 2 u^{(0)} u^{(1)} \varepsilon 
 + O(\varepsilon^2)
\\
u(x+h) & = &
u(x) + u'(x) \pas + (1/2) u''(x) \pas^2 + O(\pas^3)
\\
{u(x+ \pas) - u(x) \over \pas}
& = &
u^{(0)'} + (u^{(1)'} + (1/2) u^{(0)''}) \varepsilon + O(\varepsilon^2).
\end{eqnarray}
The equations of orders $n=0$ and $n=1$
\begin{eqnarray}
E^{(0)} & = &
u^{(0)'} + u^{(0)^2} =0
\\
E^{(1)} & = &
E^{(0)'} u^{(1)} + (1/2) u^{(0)''} = 0,\
E^{(0)'} =\partial_x + 2 u^{(0)}.
\end{eqnarray}
have the general solution
\begin{eqnarray}
u^{(0)} & = &
\chi^{-1}, \chi=x-x_0,\ x_0 \hbox{ arbitrary}
\\
u^{(1)} & = &
u_{-1}^{(1)} \chi^{-2} - \chi^{-2} \Log \psi,\ 
\psi=x-x_0,\
u_{-1}^{(1)} \hbox{ arbitrary},
\end{eqnarray}
and the movable logarithm proves the instability as soon as order $n=1$,
at the Fuchs index $i=-1$.

{\it Remark}.
The only restriction on $u^{(0)}$ is not to be what is called
a singular solution
(not obtainable from the general solution by assigning values to the
arbitrary data),
i.e.~it can be either the general solution (as above) or a particular one,
it can also be either global (as above) or local (Laurent series).

The processing of the example (\ref{eqP1CompleteDiscrete})~\cite{CM1996}
isolates three values of $\overrightarrow \lambda$, with $g=x$.

The first value $\overrightarrow \lambda=(2/3,1/3,0)$
(case $a=1$ in Ref.~\cite{GRP1991})
corresponds to the d--(P1) (\ref{eqdP1BrezinKazakov}) with a Lax pair 
found in quantum gravity, the condition is then sufficient.
The second value $(1,0,0)$ 
(case $a=0$ in Ref.~\cite{GRP1991})
corresponds to a candidate d--(P1) with a second order Lax pair
\cite{CM1996}.
The third value $(1/2,1/4,1/4)$
defines an equation equivalent to that for $(1,0,0)$ 
under a discrete birational transformation
(\ref{eqGroupBirationalDiscrete})~\cite{RG96}.

{\it Remark}.
Following Painlev\'e~\cite{PaiBSMF},
one should in fact search for the ``complete equation'',
i.e.~for all the admissible nondominant terms which can be added to
the candidate d--(P1) (\ref{eqP1CompleteDiscrete}) without destroying the PP.
According to the already known difference or $q-$difference d--(P1) candidates
(fifteen to our knowledge),
the only admissible complementary terms seem to be
\begin{eqnarray}
& &
\pas^2 X \discr(u),\
\pas^2 u \discr(u^2),\
\pas^3 (\overline{u} - \underline{u})/ (2  \pas),\
\nonumber
\\
& &
\pas^4 X^2,\
\pas^4 X \discr(u^2),\
\pas^4 u^2 \discr(u^2),
\end{eqnarray}
in which $X$ is $x$ for difference equations
or the suitable exponential function of $x$ for $q-$difference equations.

\subsection{Comparison of the two main methods}
\indent

The two main methods which define the discrete Painlev\'e test,
namely the singularity confinement method 
and the perturbation of the continuum,
happen to find the same necessary conditions
when applied to a sample of equations :
the candidate d--(P1) (\ref{eqP1CompleteDiscrete}),
a candidate discrete Chazy equation of class III not yet fully integrated
\cite{LabrunieConteChazyIII},
discrete nonlinear Schr\"odinger equations
or
various discrete Korteweg-de Vries equations
\cite{LabrunieThese}.
Other examples are currently under investigation
to detect situations where the two methods would produce complementary,
not identical results.

Since it only relies on the existence of a continuum limit,
our method can be extended without difficulty
to equations in an arbitrary number of dependent or independent variables,
whether the equations be discrete or mixed continuous and discrete.

The singularity confinement method has also been extended to such situations
\cite{RGT92,RGT93};
however, in the case of $m$ discrete independent variables,
one must check in addition 
that the result of the iteration is independent of the path followed 
on the $m-$th dimensional lattice.

In the case of equations of second degree and higher,
our method is unchanged since, again, it relies on the continuum limit,
for which this technical question is settled.
In such a case,
the confinement method must make at each step a coherent choice of 
determination then compute the confinement condition.

\section{Towards the discrete Painlev\'e and Gambier equations}
\label{sectiondPn}
\indent

The task of finding discrete analogues of the fifty canonical equations
of Gambier d--(Gn)
is, at present time, far from being achieved.
Let us give here a brief summary of the situation
and a few lines of conduct to improve it.

In the continuous case, the PP has been proved
either by explicitly linearizing,
or by integrating with elliptic functions,
or by proving the irreducibility and the absence of movable critical 
singularities.
This is equivalent to
either linearize
or find a Lax pair.

In the discrete case, after having performed the discrete Painlev\'e test
in order to isolate candidates d--(Gn),
one must do the same :
either linearize
or find a discrete Lax pair.

The only sure informations at our disposal are :
the fifty continuous (Gn) equations of course,
the exact discrete elliptic equations.

To be precise, 
according to the conjecture of Section \ref{sectionBasicRules},
one should look for d--(Gn) equations satisfying the following criteria.

\begin{enumerate}

\item
Each d--(Gn) passes the discrete Painlev\'e test.

\item
Each d--(Gn) has, like (Gn), order two and degree one.

\item
Each d--(Gn) 
must either be explicitly linearizable,
or possess a first integral identical to an elliptic equation,
or possess a matricial Lax pair $(A,B,z)$
admitting as continuum limit a Lax pair $(L,M,t)$ of (Gn).
This implies an order two for these discrete Lax pairs.

\item
The d--(Pn) equations
satisfy a confluence cascade 
admitting as continuum limit the cascade of Painlev\'e and Gambier
\cite{PaiCRAS1906,GambierThese}
down to the Weierstrass level included,
see formulae below.

\end{enumerate}

As an example of some difficulties,
consider the (G27) equation in the particular case of Ermakov and Pinney
\begin{eqnarray}
& &
-u u'' + (1/2) u'^2 + f(x) u^2 - c^2/2 =0,\
c \not=0,\
c \hbox{ constant},
\label{eqErPiContinu}
\end{eqnarray}
which is linearizable
either by derivation
\begin{eqnarray}
& &
-u''' + 2 f u' + f' u =0,
\label{eqErPiTransfo1Continu}
\end{eqnarray}
or by the singular part transformation
\begin{eqnarray}
& &
u^{-1}=c (\Log(\psi_2 / \psi_1))',\
\psi_k'' - (f/2) \psi_k=0,\ k=1,2.
\label{eqErPiTransfo2Continu}
\end{eqnarray}

Three discrete candidates for (\ref{eqErPiContinu})
have been proposed,
the first one of degree one~\cite{CHM96}
\begin{eqnarray}
& &
-u (\overline{u} - 2 u + \underline{u}) \pas^{-2}
+(1/2) (\overline{u} - u) (u - \underline{u}) \pas^{-2}
+ f    (\overline{u} + u) (u + \underline{u})/4
-c^2/2 =0,
\end{eqnarray}
the second one also of degree one~\cite{GR96}
\begin{eqnarray}
& &
-u (\overline{u} - 2 u + \underline{u}) \pas^{-2}
+(1/2) (\overline{u} - u) (u - \underline{u}) \pas^{-2}
+ f  u \underline{u} 
-c^2/2
\nonumber
\\
& &
+(\pas /2) c (\overline{u} - 2 u + \underline{u}) \pas^{-2}
=0,
\end{eqnarray}
the third one of degree two~\cite{CHM96,CHM97}
\begin{eqnarray}
& &
\left(
(\overline{u} - \underline{u} - (2 + \pas^2 f/2)^2 u)^2
-(2 + \pas^2 f/2)^2 (4 u \underline{u} + \pas^2 c^2)
\right) / (8 \pas^2)
=0.
\end{eqnarray}

The third one has been linearized by discrete analogues of {\it both}
transformations
(\ref{eqErPiTransfo1Continu}) and (\ref{eqErPiTransfo2Continu})
\cite{CHM97}
but it has degree two.
The first one has been linearized by a discrete analogue of
(\ref{eqErPiTransfo1Continu}) only,
but it has two nice features :
it obeys the rules and has no complementary terms.
The second one, to our knowledge, has not yet been integrated.

Before proceeding,
let us recall for later use the definition of (P4)
\begin{eqnarray}
u''
&=&
{u'^2 \over 2 u} + {3 \over 2} u^3 + 4 x u^2 + 2 (x^2 - \alpha) u 
+ {\beta \over u}
\end{eqnarray}
and the restriction to the subset ((P4), (P2), (P1), $\wp$)
of the confluence cascade from
(Pn)$(x,u,\alpha,\beta,\gamma,\delta)$ to
(Pm)$(X,U,A,B,C,D)$, $m<n$,
\begin{eqnarray}
4 \to 2 : 
& &
(x,u,\alpha,\beta)
=
(\varepsilon X - \varepsilon^{-3}/4,
\varepsilon^{-3}/4 + \varepsilon^{-1} U,
- \varepsilon^{-6}/32 - 2 A,
- \varepsilon^{-12}/512),
\label{eqCoalesce42}
\\
2 \to 1 : 
& &
(x,u,\alpha)
=
(-6 \varepsilon^{-10} + \varepsilon^2 X,
\varepsilon^{-5} + \varepsilon U,
4 \varepsilon^{-15}),
\label{eqCoalesce21}
\\
1 \to \wp : 
& &
(x,u)
=
(-\varepsilon^{-4} g_2/2 + \varepsilon X, \varepsilon^{-2} U),
\label{eqCoalesce1wp}
\end{eqnarray}
with $\varepsilon \to 0$.

Let us from now on restrict to discretizations without complementary terms,
such as, for d--(P1), equation (\ref{eqP1CompleteDiscrete})
but not equation (\ref{dP1ordre4}).

For d--(P1),
three candidates (\ref{eqP1CompleteDiscrete}) pass the test :
\hfill \break \noindent
$\overrightarrow \lambda=(2/3,1/3,0), (1,0,0), (1/2,1/4,1/4)$.
Among these,
the first one is the only one to admit a confluence to the discrete
equation of Weierstrass and,
since it satisfies the other above criteria,
it definitely can be called 
``the (unique) d--(P1) without complementary terms''.

For d--(P2),
this is (\ref{eqdP2PeriwalShevitz}) the good equation,
because of both its Lax pair 
and the confluence from
d--(P2)$(u,x,\pas,\alpha)$ to d--(P1)$(U,X,\Pas)$~\cite{RGH1991};
this confluence is better written as
\def\TEMP{\lambda}
\begin{equation}
(x,\pas,u,\alpha)
=
(\TEMP^{2} (X - 6 \varepsilon^{-12}),
 \TEMP^{2} \Pas,
 \TEMP (\varepsilon^{-6} + U),
 4 \TEMP^{3} \varepsilon^{-18}),\
\TEMP^{-6}= \varepsilon^{-6} + \Pas^2 \varepsilon^{-12},
\label{eqCoalesce21discret}
\end{equation}
which proves that its continuum limit
is the continuous confluence (\ref{eqCoalesce21}).

For the four other equations (P3), (P4), (P5), (P6),
there is not yet a fully satisfactory result,
i.e.~for each of them at least one of the enumerated points is not satisfied.

Two discrete (P3) candidates have been proposed.
The $q-$(P3) candidate~\cite{RGH1991}
has a matricial Lax pair~\cite{PNGR1992} of order four, not two,
without clear continuum limit.
The d--(P3) candidate,
system (25ab) of two equations in two fields in Ref.~\cite{GNPRS1994},
has a fourth order Lax pair~\cite{GNPRS1994,JS1996},
but it also has a subtle drawback.
Indeed,
after elimination of the second field,
the discrete equation has degree two,
although its continuum limit (P3) has degree one.
The second order Lax pair (\ref{eqLaxdP3Begin})
is a starting point to remedy this situation.

The d--(P4) candidate~\cite{RGH1991}
\begin{eqnarray}
& &
E \equiv
-u (\overline{u} - 2 u + \underline{u}) \pas^{-2}
+(1/2) (\overline{u} - u) (u - \underline{u}) \pas^{-2}
\label{eqdP4}
\\
& &
+ u^2  (u \overline{u} + u \underline{u} + \overline{u} \underline{u})/2
+ (x u + x^2/2) (\overline{u} + u) (u + \underline{u})
- 2 \alpha u^2 + \beta=0
\nonumber
\end{eqnarray}
admits a confluence to the d--(P2)~\cite{RGH1991},
and one can even check that this
confluence is independent of the stepsize and given by (\ref{eqCoalesce42}).
Although this is evidently the good d--(P4),
its Lax pair is still unknown.

No d--(P5) candidate is known.
The $q-$(P5) candidate~\cite{RGH1991} has the correct degree (one),
it admits confluences~\cite{RGH1991} to both the $q-$(P3) candidate
and the d--(P2),
but its Lax pair is unknown.

Two discrete (P6) candidates have also been proposed.
Both the $q-$(P6) candidate~\cite{JS1996} 
and  the  d--(P6) candidate~\cite{RGO1996}
are defined as a system of two equations in two fields
and, just like for the d--(P3) candidate,
the discrete equation in a single field has second order but also
second degree (as remarked in Ref.~\cite{JS1996}),
although its continuum limit is (P6) itself.
The d--(P6) candidate has not yet a Lax pair,
but the $q-$(P6) candidate has one by construction,
since Jimbo and Sakai started in fact from a $q-$difference isomonodromic
deformation problem in order to obtain their system.

\section{Conclusion
\label{sectionConclusion}}
\indent

These precise definitions and guidelines may render
more systematic the search for discrete analogues of differential equations
integrable in the sense of Painlev\'e.

The most fundamental anchor point consists of the discrete elliptic equations
because they are exact.

As to the new perturbative method for the discrete Painlev\'e test,
its full applicability 
(to any number of independent variables, whether discrete or
mixed discrete-continuous)
should make it a quite efficient tool to investigate which discretizations
of partial differential equations
preserve the Painlev\'e property.

{\it Acknowledgements}.
\indent

We thank Alan K.~Common for having drawn our attention on the work of Potts,
and J.-M.~Drouffe for the use of his efficient computer algebra language
AMP~\cite{Drouffe}.

Both authors acknowledge the financial support of the Tournesol grant
T 95/004.
M.~M.~acknowledges the financial support extended by 
Flanders's Federale Diensten voor Wetenschappelijke, Technische en Culturele
Aangelegenheden in the framework of the IUAP III no.~9. 

\vfill \eject


\end{document}